\documentclass{PoS}

\usepackage[italian,english]{babel}
\usepackage{amssymb}
\usepackage{amsfonts}
\usepackage{amsmath}
\usepackage{fancyhdr}
\usepackage{lineno}
\usepackage{subfigure}
\newcommand{\nn}{\nonumber}
\newcommand{\lsim}{\mathrel{\mathop{\kern 0pt \rlap
  {\raise.2ex\hbox{$<$}}}
  \lower.9ex\hbox{\kern-.190em $\sim$}}}
\newcommand{\gsim}{\mathrel{\mathop{\kern 0pt \rlap
  {\raise.2ex\hbox{$>$}}}
  \lower.9ex\hbox{\kern-.190em $\sim$}}}

\newcommand{\be}{\begin{equation}}
\newcommand{\ee}{\end{equation}}
\newcommand{\bea}{\begin{eqnarray}}
\newcommand{\eea}{\end{eqnarray}}

\def\etmiss{\not\!\!{E_T}}
\def\ptmiss{\not\!\!{p_T}}

\usepackage{pstricks}
\usepackage{color}
\usepackage{multirow}
\usepackage{cite}

\title{Charged Higgs bosons in the extended supersymmetric scenario at the LHC}

\ShortTitle{Charged Higgs bosons in the extended supersymmetric scenario at the LHC}

\author{\speaker{Priyotosh Bandyopadhyay}%
\\
        Dipartimento di Matematica e Fisica "Ennio De Giorgi", \\ Universit\`a del Salento and INFN-Lecce,  Via Arnesano, 73100 Lecce, Italy\\and \\
        Indian Institute of Technology Hyderabad, Kandi,  Sangareddy-502287, Telengana, India\\
        E-mail: \email{priyotosh.bandyopadhyay@le.infn.it, bpriyo@iith.ac.in}}

\abstract{We investigate an extension of the Minimal
Supersymmetric Standard Model (MSSM) containing a $SU(2)$ Higgs triplet of zero hypercharge and a gauge singlet.  We focus on a scenario of this model which allows a light pseudoscalar and/or a scalar
below $100$ GeV in the spectrum, consistent with the most recent data from the LHC and the earlier data from the LEP experiments. The triplet extension gives rise to two additional charged Higgs bosons which do not couple to fermions but can decay into $Z W^\pm$. The presence of a very light pseudoscalar  opens $a_1 W^\pm$ decay mode for the light charged Higgs boson just like $Z_3$ symmetric singlet-extension (NMSSM). A triplet type charged Higgs can be produced via vector boson fusion at the tree-level which is absent in 2HDM and MSSM and NMSSM. If such charged Higgs boson is pair produced both $Z W^\pm$ and $a_1 W^\pm$ decay modes can be probed which can prove  the existence of triplet and singlet both.  }

\FullConference{Fourth Annual Large Hadron Collider Physics\\
		 13-18 June 2016\\
		 Lund, Sweden}

\begin{document}

\section{Introduction}
LHC has discovered one Higgs boson with mass around 125 GeV and thus proved the role of at least one 
scalar in electro-weak symmetry breaking (EWSB) \cite{CMS, ATLAS}.  However the question of role another scalar  in EWSB still remains.  Though Standard Model (SM) is successful in explaining the electro-magnetic, weak and strong interactions in particle physics, it has its own caveats. First of all it does not have a suitable cold dark matter (DM) candidate that can explain the current DM relic abundance. Secondly 
the neutrino mass and their tinniness cannot be explained without fine-tuning. Thirdly and one of the most important problem inside SM is that the only scalar in SM, the Higgs boson's mass is not protected by any symmetry. This leads to the famous gauge-hierarchy problem in SM, where Higgs mass diverges 
quadratically to the scale of the theory due to quantum corrections.  There are many other anomalies and caveats inside SM and so numerous theories beyond SM have been proposed in order to rectify some of those problems. Supersymmetry is one of the most popular scenarios where it not only solves gauge hierarchy problem of the Higgs mass but also provides a good DM candidate with correct range of mass and annihilation cross-section required to attain the observed DM relic.  However minimal form of the theory suffers a fine-tuning problem since the Higgs discovery with mass around 125 GeV.  In the minimal supersymmetric extension of SM (MSSM) the lightest Higgs mass is $\lesssim m_Z$ at tree-level and to achieve Higgs boson mass around $\sim 125$ GeV one either needs large supersymmetric (SUSY) mass scale and/or large mass-splitting between the mass values of the super-partners of the top quarks \cite{carena}. Thus below TeV scale SUSY is rather demotivated in these minimal scenarios.

Introduction of additional scalar not only solve this apparent fine-tuning problem but also justifies the theoretical possibilities of the Higgs bosons of other kind; namely the Higgs bosons from different representation of $SU(2)\times U(1)$. Additional Higgs bosons either can from singlet or triplet representations of $SU(2)$ which can take part in the EWSB. These Higgs bosons not only contribute to the SM Higgs boson mass at the tree-level but also at quantum level which reduce the required quantum correction from the strong sector. Introduction of additional doublet  and/or triplet that take part in EWSB, i.e., the neutral parts that get the vacuum expectation value (vev), give rise to a charged scalar as a physical excitation. Thus the finding a charged Higgs boson will be direct of the existence of other Higgs doublet or triplet which takes part in the EWSB. 

In this article we discuss the different possibilities of charged Higgs bosons that can appear in the extended Higgs sectors with supersymmetry. The minimal extension with additional Higgs doublet (MSSM) gives rise to a single doublet type charged Higgs boson, where as extension with a $SU(2)$ triplet can give rise to two or more triplet type charged Higgs bosons even with doubly charged possibility. The experimental searches for light charged Higgs boson are mostly biased toward the doublet type charged Higgs which decays via fermionic modes, i.e., $\tau \,\nu$ and/or $t\, b$ \cite{ChCMS,ChATLAS}. Here we consider the extension of the Higgs sector with $Y=0$ triplet and  a SM gauge singlet in the context of supersymmetry. The existence of the additional singlet scalar opens up the possibility of the
a very light pseudo-scalar (hidden/buried) which can be few GeV to few tens of GeV. Such light pseudoscalar makes the charged Higgs phenomenology even more interesting by introducing new decay modes. 

In section~\ref{model} we describe the model briefly and discuss the Higgs boson mass spectrum. In section~\ref{chpheno} the charged Higgs phenomenology for the triplet type charged Higgs is discussed. 
Section~\ref{chnmssm} is devoted on the non-standard decays of the charged Higgs boson in $Z_3$ NMSSM \cite{Ellwanger}.  In section~\ref{dis} we discuss how different such extensions can be distinguished at colliders and conclude.
\section{The model}
\label{model}
The superpotential of the triplet and singlet extended supersymmetric SM (TNMSSM) \cite{TNMSSM1, TNMSSM2, TNMSSM3}, $W_{TNMSSM}$, contains a SU(2) triplet $\hat{T}$ of zero hypercharge ($Y=0$)  together with a SM gauge singlet ${\hat S}$, added to the superfield content of the MSSM,

\begin{equation}
 W_{TNMSSM}=W_{MSSM} + W_{TS}.
 \end{equation}
The structure of its triplet and singlet extended superpotential can be written as 

 \begin{equation}
W_{TS}=\lambda_T  \hat H_d \cdot \hat T  \hat H_u\, + \, \lambda_S \hat S\,  \hat H_d \cdot  \hat H_u\,+ \frac{\kappa}{3}\hat S^3\,+\,\lambda_{TS} \hat S \, \textrm{Tr}[\hat T^2] .
\label{spt}
 \end{equation}
 
 The triplet and doublet superfields are given by 
\begin{equation}\label{spf}
 \hat T=\begin{pmatrix}
       \sqrt{\frac{1}{2}}\hat T^0 & \hat T_2^+ \cr
      \hat T_1^- & -\sqrt{\frac{1}{2}}\hat T^0
       \end{pmatrix},\, \hat{H}_u= \begin{pmatrix}
      \hat H_u^+  \cr
       \hat H^0_u
       \end{pmatrix},\, \hat{H}_d= \begin{pmatrix}
      \hat H_d^0  \cr
       \hat H^-_d
       \end{pmatrix}.
 \end{equation}
 
 Here $\hat T^0$ denotes a complex neutral superfield, while  $\hat T_1^-$ and $\hat T_2^+$ are the charged Higgs superfields.  The MSSM Higgs doublets are the only superfields which couple to the fermion multiplet via Yukawa coupling. All the coefficients involved in the Higgs sector are chosen to be real in order to preserve CP invariance. The breaking of the $SU(2)_L\times U(1)_Y$ electroweak symmetry is then obtained by giving real vevs to the neutral components of the Higgs field
 \be
 <H^0_u>=\frac{v_u}{\sqrt{2}}, \, \quad \, <H^0_d>=\frac{v_d}{\sqrt{2}}, \quad <S>=\frac{v_S}{\sqrt{2}}, \, \quad\, <T^0>=\frac{v_T}{\sqrt{2}},
 \ee
 which give mass to the $W^\pm$ and $Z$ bosons
 \be
 m^2_W=\frac{1}{4}g^2_L(v^2 + 4v^2_T), \, \quad\ m^2_Z=\frac{1}{4}(g^2_L \, +\, g^2_Y)v^2, \, \quad v^2=(v^2_u\, +\, v^2_d), 
\quad\tan\beta=\frac{v_u}{v_d} \ee
 and also induce, as mentioned above, a $\mu$-term of the form $ \mu_D=\frac{\lambda_S}{\sqrt 2} v_S+ \frac{\lambda_T}{2} v_T$. The triplet vev $v_T$ is strongly  constrained by the global fit to the measured value of the $\rho$ parameter \cite{rho}  which restricts its value to $v_T \leq 5$ GeV  and in the numerical analysis we have chosen $v_T =3$ GeV.

In the TNMSSM, the neutral CP-even mass matrix is $4$-by-$4$, since the mixing terms involve the two $SU(2)$ Higgs doublets, the scalar singlet $S$ and the neutral component of the Higgs triplet. Being the Lagrangian CP-symmetric, we are left with four CP-even, three CP-odd  and three charged Higgs bosons as shown below
 \bea\label{hspc}
  \rm{CP-even} &&\quad \quad  \rm{CP-odd} \quad\quad   \rm{charged}\nn \\
 h_1, h_2, h_3, h_4 &&\quad \quad a_1, a_2, a_3\quad \quad h^\pm_1, h^\pm_2, h^\pm_3. 
 \eea
The neutral Higgs bosons are combination of doublets, triplet and singlets, whereas the charged Higgs bosons are a combination of doublets and triplet only. We will denote with $m_{h_i}$ the corresponding mass eigenvalues, assuming that one of them will coincide with the 125 GeV Higgs $(h_{125})$ boson detected at the LHC. We investigate the scenario where one 
(or more) scalar or pseudoscalar with a mass $< 125$ GeV is allowed, which we call a {\em hidden Higgs} scenario. 

At tree-level the maximum value of the lightest neutral Higgs has additional contributions from the triplet and the singlet sectors respectively. The numerical value of the upper bound on the lightest CP-even Higgs can be extracted from the relation
\be\label{hbnd}
m^2_{h_1}\leq m^2_Z(\cos^2{2\beta} \, +\, \frac{\lambda^2_T}{g^2_L\,+\,g^2_Y }\sin^2{2\beta}\, +\, \frac{2\lambda^2_S}{g^2_L\,+\,g^2_Y }\sin^2{2\beta}),
\ee
which is affected on its right-hand-side by two additional contributions from the triplet and the singlet. These can raise the allowed tree-level Higgs mass. Both contributions are proportional to $\sin{2\beta}$, and thus they can be large for a low value of $\tan{\beta}$. The additional contributions coming from the triplet and the singlet reduce the fine-tuning of the supersymmetric mass scale 
required to attain the lightest CP-even Higgs boson mass of 125 GeV. These extra contributions at the tree-level are large for a low $\tan{\beta}$, and so we do not need radiative corrections in order to match the observed mass. However for a large $\tan{\beta}$ value, these extra scalars contribute enough at higher orders, reducing the required radiative corrections from the squarks and the corresponding supersymmetric mass scale. Thus $\lesssim$ TeV scale SUSY can still be realised in the extended supersymmetric sectors \cite{TNMSSM1, pbas1}. 

\section{Charged Higgs bosons in TNSSM}\label{chpheno}
As already discussed before and in \cite{TNMSSM1,TNMSSM2}, there are four CP-even neutral ($h_1,h_2,h_3,h_4$), three CP-odd neutral ($a_1,a_2, a_3$) and three charged Higgs bosons ($h_1^\pm,h_2^\pm,h_3^\pm$). In general the mass eigenstates are obtained via a mixing of the two Higgs doublets, the triplet and the singlet scalar. However, the singlet does not contribute to the charged Higgs bosons. The charged Higgs  bosons are mixed states generated only by the $SU(2)$ doublets and triplets and the rotation from gauge eigenstates are defined as Eq.~\ref{chmix}
\bea\label{chmix}
h_i= \mathcal{R}^S_{ij} H_j, \quad 
a_i= \mathcal{R}^P_{ij} A_j, \quad 
h^\pm_i= \mathcal{R}^C_{ij} H^\pm_j\nn
\eea
where the eigenstates on the left-hand side are mass eigenstates whereas the eigenstates on th right-hand side are gauge eigenstates. Explicitly we have $h_i=(h_1,h_2,h_3,h_4)$, $H_i=(H^0_{u,r},H^0_{d,r},S_r,T^0_r)$, $a_i=(a_0,a_1,a_2,a_3)$, $A_i=(H^0_{u,i},H^0_{d,i},S_i,T^0_i)$, $h_i^\pm=(h_0^\pm,h_1^\pm,h_2^\pm,h_3^\pm)$ and $H_i^+=(H_u^+,T_2^+,H_d^{-*},T_1^{-*})$. { Using these definitions we can write the doublet, singlet and triplet fraction for the scalar and pseudoscalar Higgs bosons as
\bea
h_i|_{D}=(\mathcal{R}^S_{i,1})^2+(\mathcal{R}^S_{i,2})^2, \,\, a_i|_{D}=(\mathcal{R}^P_{i,1})^2+(\mathcal{R}^P_{i,2})^2\\
h_i|_{S}=(\mathcal{R}^S_{i3})^2, \,\, a_i|_{S}=(\mathcal{R}^P_{i3})^2\\
h_i|_T=(\mathcal{R}^S_{i4})^2, \,\, a_i|_T=(\mathcal{R}^P_{i4})^2
\eea
and the triplet and doublet fraction of the charged Higgs bosons as
\bea
h_i^\pm|_D=(\mathcal{R}^C_{i1})^2+(\mathcal{R}^C_{i3})^2, \,\, h_i^\pm|_T=(\mathcal{R}^C_{i2})^2+(\mathcal{R}^C_{i4})^2 .
\eea

We call a scalar(pseudoscalar) Higgs boson doublet-like if $h_i|_D(a_i|_D)\geq\,90\%$, singlet-like if $h_i|_S(a_i|_S)\geq\,90\%$ and triplet-like if $h_i|_T(a_i|_T)\geq\,90\%$. Similarly a charged Higgs boson will be doublet-like if $h_i^\pm|_D\geq\,90\%$ or triplet-like if $h_i^\pm|_T\geq\,90\%$.}

\subsection{$h_i^\pm  \to W^\pm Z$}
The charged sector of a theory with scalar triplet(s) is very interesting due to the tree-level interactions $h_i^\pm-W^\mp-Z$ for $Y=0, \pm 2$ hypercharge triplets which break the custodial symmetry \cite{pbas3,EspinosaQuiros,tnssm, tnssma}. In the TNMSSM this coupling is given by 
\bea\label{zwch}
g_{h_i^\pm W^\mp Z}&=&-\frac{i}{2}\left(g_L\, g_Y\left(v_u\sin\beta\,\mathcal R^C_{i1}-v_d\cos\beta\,\mathcal R^C_{i3}\right)\right.\nn\\
&&\left.+\sqrt2\,g_L^2v_T\left(\mathcal R^C_{i2}+\mathcal R^C_{i4}\right)\right),
\eea
where the rotation angles are defined in Eq.~\ref{chmix}. The on-shell decay width is given by
\bea\label{chzw}
\Gamma_{h_i^\pm\rightarrow W^\pm Z}&=&\frac{G_F\,\cos^2\theta_W}{8\sqrt2\pi}m^3_{h_i^\pm}|g_{h_i^\pm W^\mp Z}|^2\nn\\
&&\times\sqrt{\lambda(1,x_W,x_Z)}\left(8\,x_W\,x_Z+(1-x_W-x_Z)^2\right)\nn\\
\eea
where $\lambda(x,y,z)=(x-y-z)^2-4\,y\,z$ and $x_{Z,W}=\frac{m^2_{Z,W}}{m^2_{h_i^\pm}}$ \cite{Asakawa}.
\begin{figure}[thb]
\begin{center}
\mbox{\subfigure[]{\includegraphics[width=0.35\linewidth]{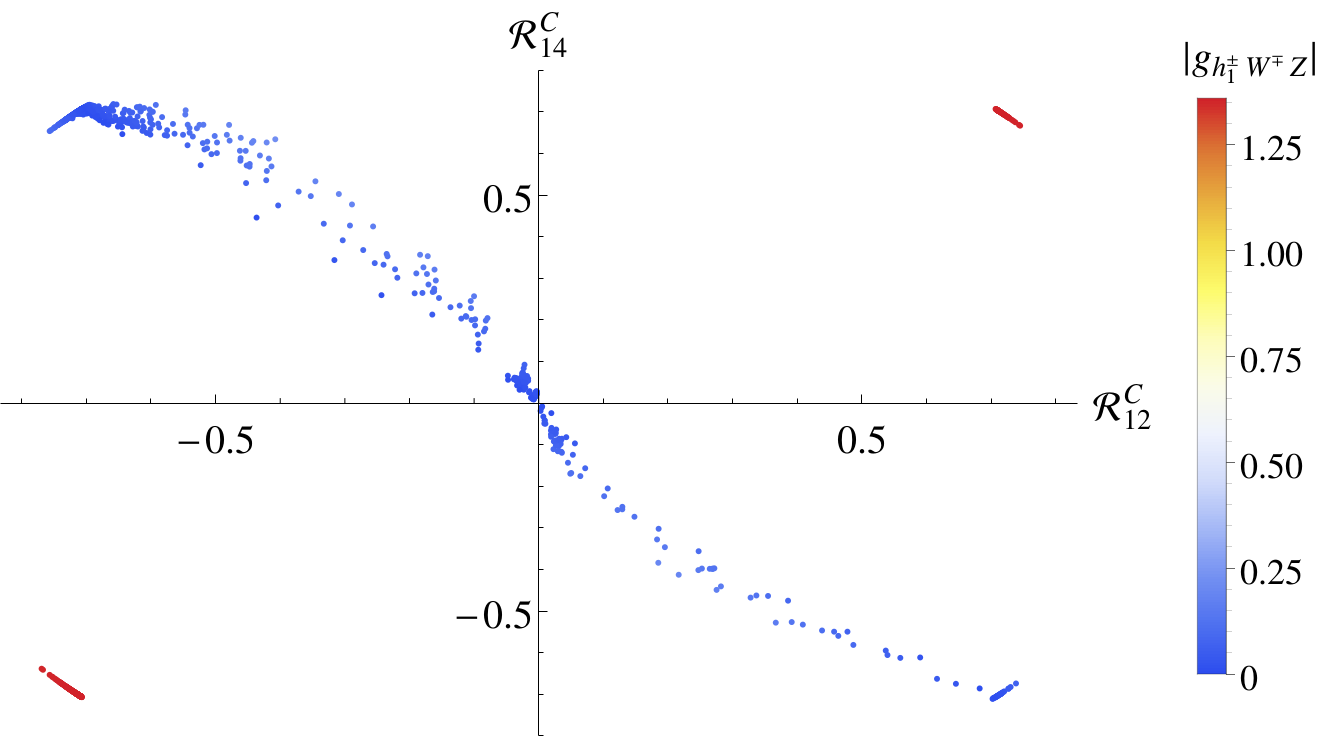}}
\subfigure[]{\includegraphics[width=.35\linewidth]{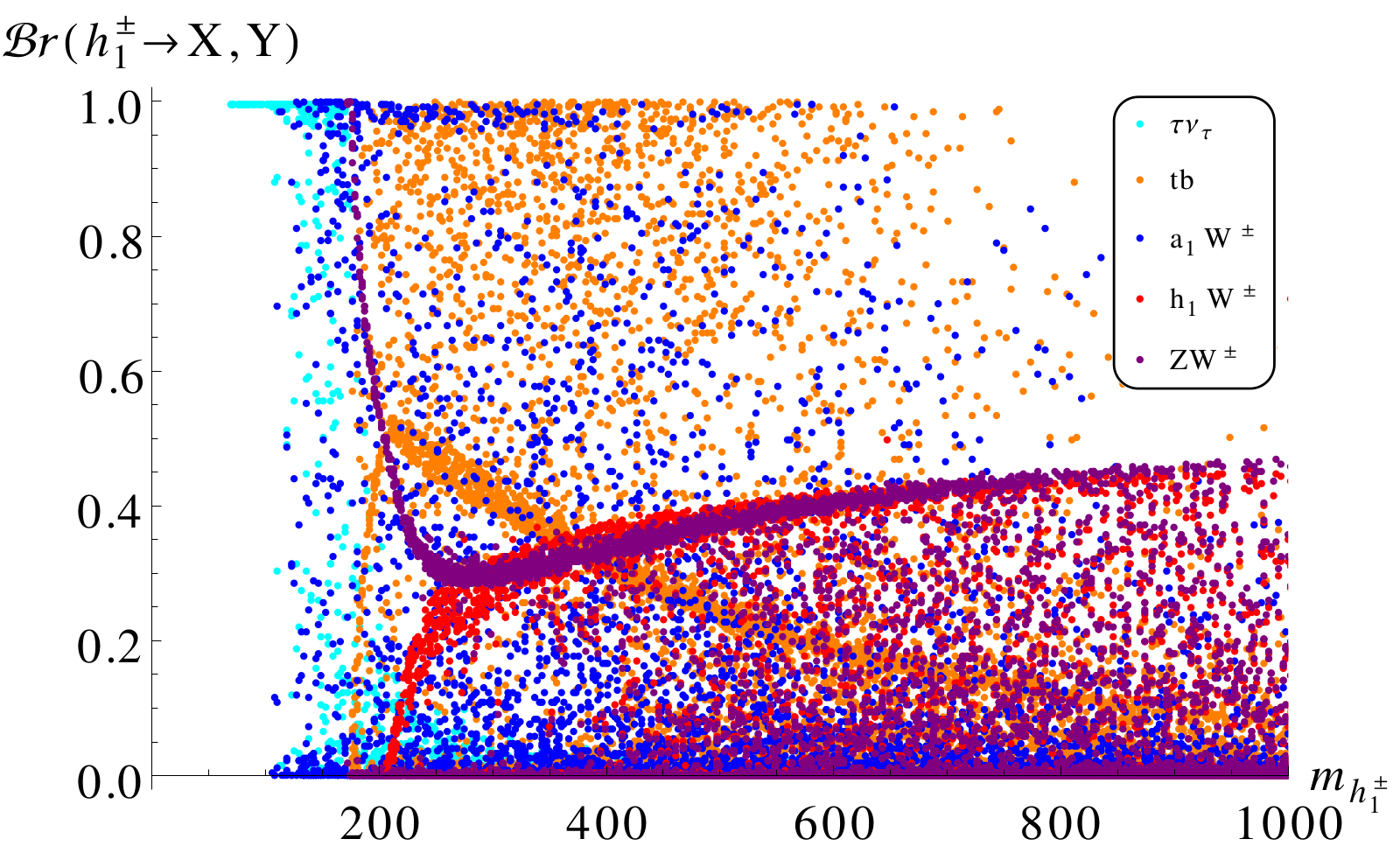}}}
\caption{Correlation of $g_{h^\pm_1 W^\mp Z}$ with  $\mathcal{R}^C_{12}$ and  $\mathcal{R}^C_{14}$ (a) and two-body decay branching fractions (b) \cite{TNMSSM3}.}\label{ghmp1WZ}
\end{center}
\end{figure}

Figure~\ref{ghmp1WZ} shows the dependency of $g_{h_i^\pm W^\mp Z}$ with respect to $\mathcal{R}^C_{12}$ and $\mathcal{R}^C_{14}$. We see that for 
$\lambda_T \sim 0$ $\mathcal{R}^C_{12}$ and $\mathcal{R}^C_{14}$ take the same sign, and hence the $h_i^\pm-W^\mp-Z$  coupling is enhanced.

\subsection{Vector boson fusion}
Neutral Higgs boson production via vector boson fusion is second most dominant 
production mode in SM. Even in 2HDM or MSSM this production mode of the neutral Higgs boson is one of the leading ones. However no such channel exist for charged Higgs boson as $h^\pm_i-W^\mp-Z$ vertex is zero at the tree-level, as long as custodial symmetry 
is preserved. The introduction of a $Y=0$ triplet breaks the custodial symmetry at tree-level, giving a non-zero $h^\pm_i-W^\mp-Z$ vertex, as shown in Eq.~\ref{zwch}. This vertex gives rise to the striking production channel of the vector boson fusion into a single charged Higgs boson, which is absent in the MSSM and in the 2-Higgs-doublet model (2HDM) at tree-level. This is a signature of the triplets with $Y=0, \pm 2$ which break custodial symmetry at the tree-level.
\begin{figure}[t]
\begin{center}
\mbox{\subfigure[]{\includegraphics[width=0.25\linewidth]{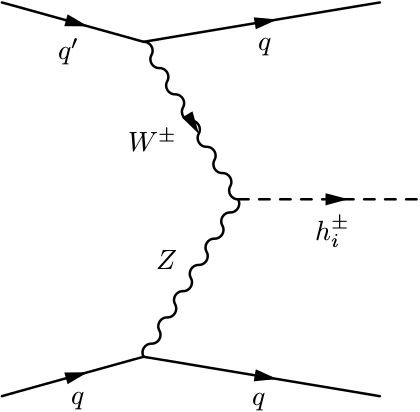}}
\hskip 25 pts
\subfigure[]{\includegraphics[width=0.45\linewidth]{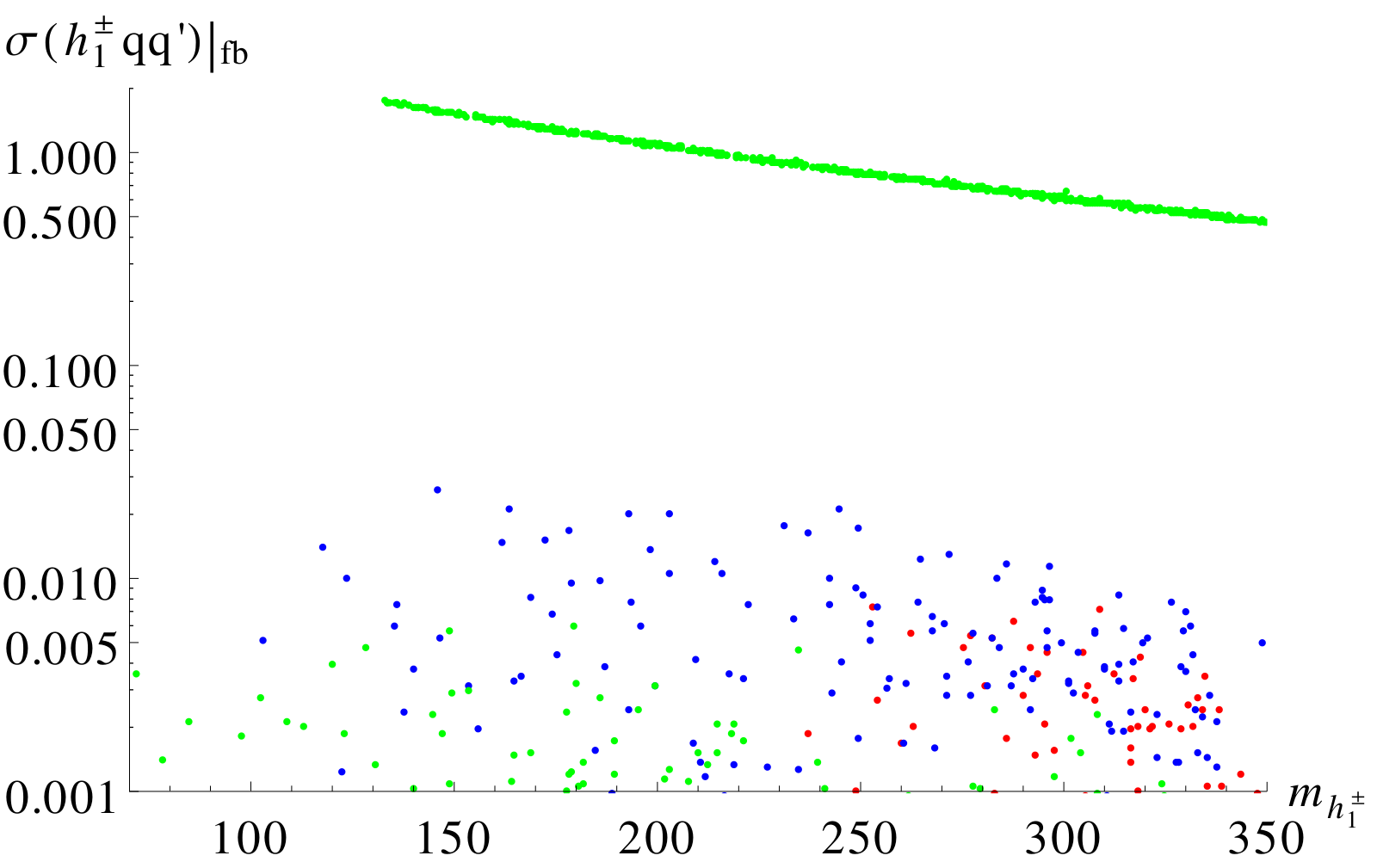}}}
\caption{The Feynman diagram for the charged Higgs production via vector boson fusion at the LHC (a)
and the production cross-section of a light charged Higgs boson via  vector boson fusion versus the light charged Higgs boson mass $m_{h^\pm_1}$ (b) \cite{TNMSSM3}.}\label{prodvvfch}
\end{center}
\end{figure}

Figure~\ref{prodvvfch}(a) shows the Feynman diagram for the charged Higgs production via vector boson fusion at the LHC and Figure~\ref{prodvvfch}(b) presents the cross-section variation with respect to the lightest charged Higgs boson mass $m_{h^\pm_1}$. As expected, doublet-like points (in red) have very small cross-sections, and for the mixed points (in blue) we see a little enhancement. Green points describe the cross-sections for the triplet-like points. We see that a triplet-like charged Higgs boson does not necessarily guarantee large values for the cross-section. As one can notice from Eq.~\ref{zwch}, the coupling $g_{h_1^\pm W^\mp Z}$ is a function of $\mathcal{R}^{C}_{12}$ and $\mathcal{R}^{C}_{14}$ and their relative sign plays an important role. From Figure~\ref{ghmp1WZ} we see that only in the decoupling limit, where where $\lambda_T=0$, both $\mathcal{R}^{C}_{12}$ and $\mathcal{R}^{C}_{14}$ take the same sign, thereby enhancing the $h_1^\pm- W^\mp -Z$ coupling and thus the cross-section.  It can been seen that only for lighter masses $\sim 150-200$ GeV the cross-sections is around few fb. Such triplet-like charged Higgs bosons can be probed at the LHC as a single charged Higgs production channel without  the top quark. This channel thus can be used to distinguish from other known single charged Higgs production mode in association with the top quark, which characterises  a doublet-like charged Higgs boson.

\subsection{Charged Higgs pair production }
\begin{figure}[thb]
\begin{center}
\mbox{\subfigure[]{
\includegraphics[width=0.4\linewidth]{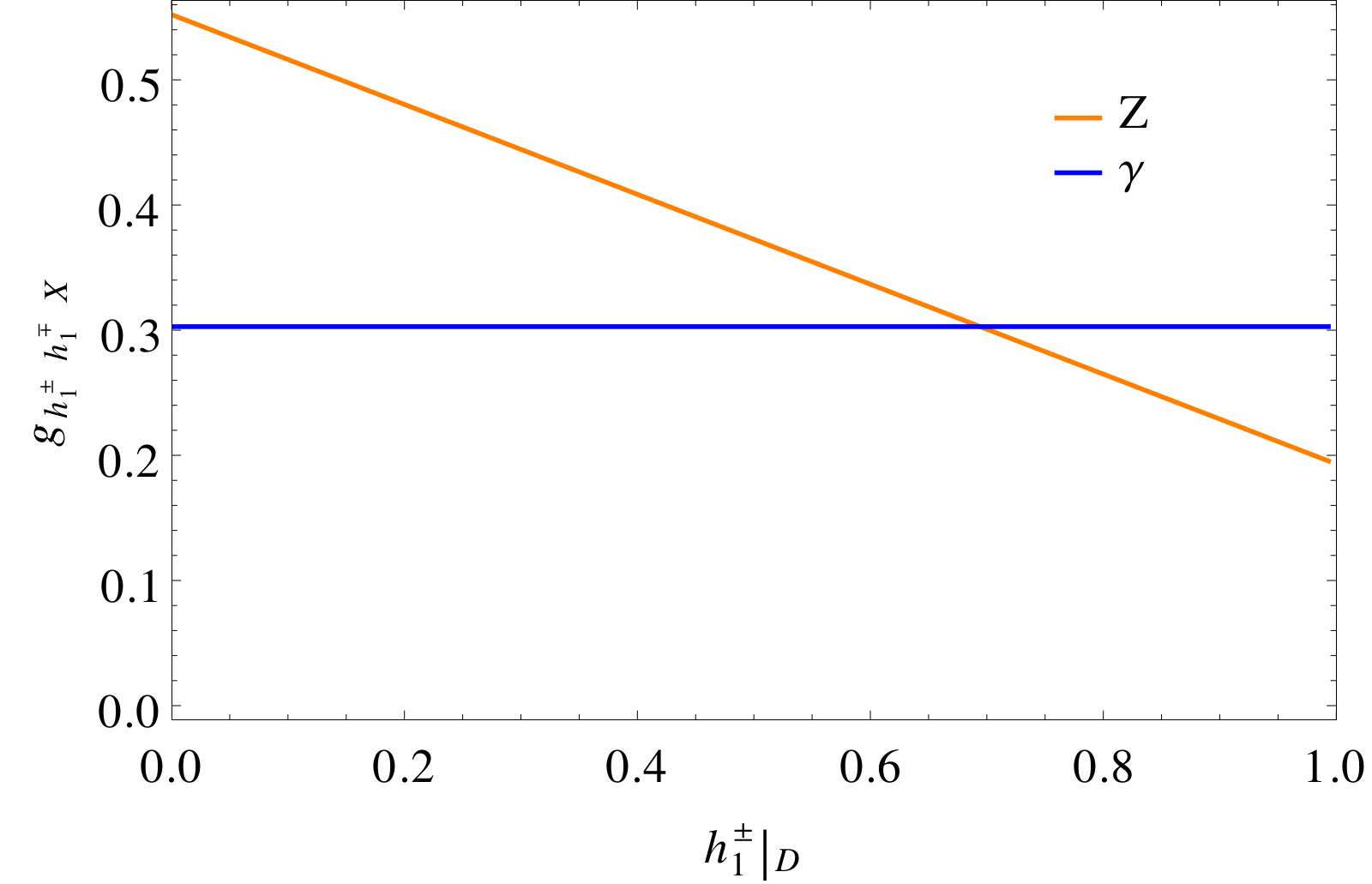}}
\subfigure[]{\includegraphics[width=0.4\linewidth]{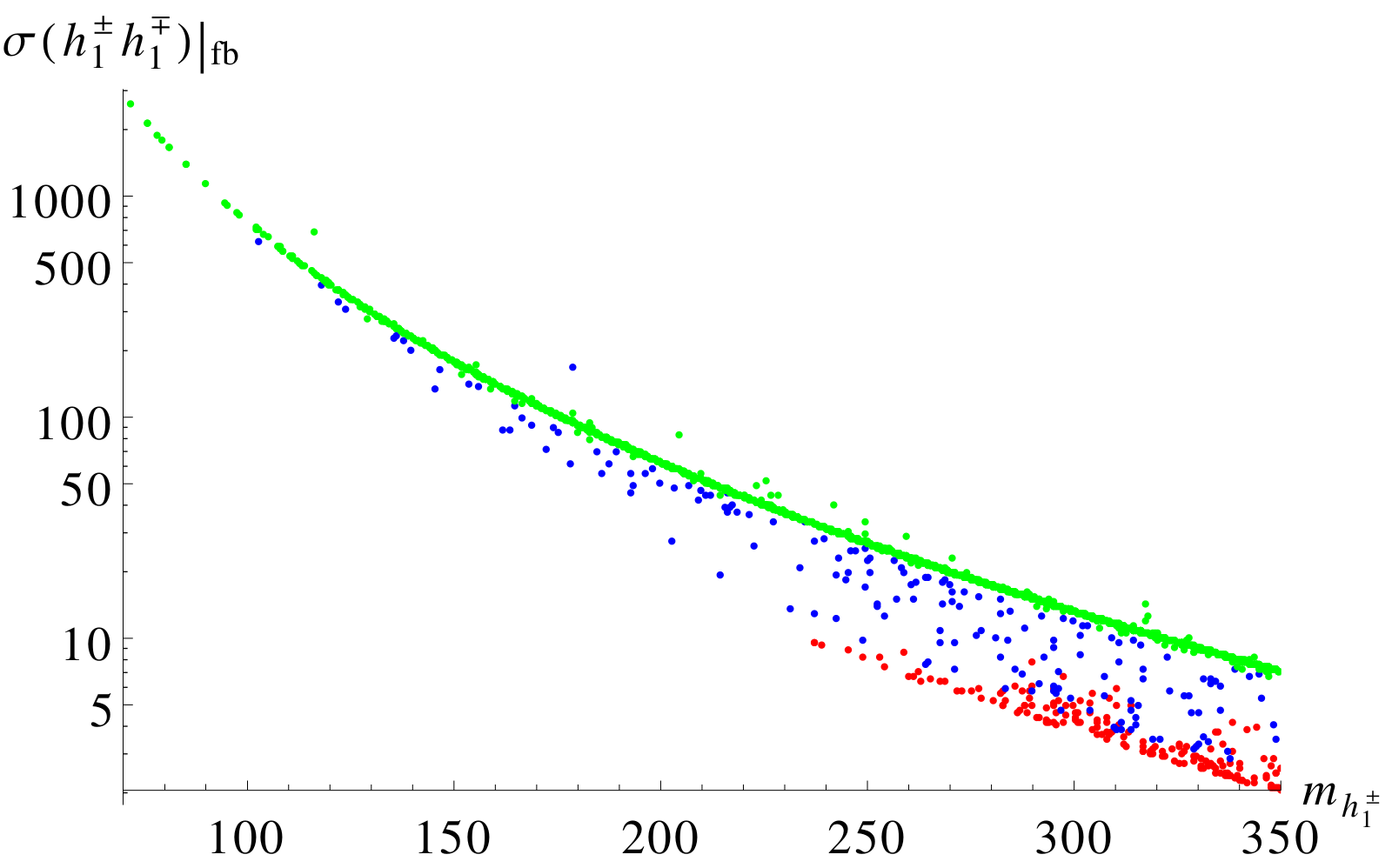}}}
\caption{ $g_{h_1^\pm h_1^\mp X}$ as a function of the doublet fraction of the lightest charged Higgs boson (a). The production cross-section of  $h^\pm_1h^\mp_1$ versus  $m_{h^\pm_1}$ (b) \cite{TNMSSM3}.}\label{ZphoHpmHpm}
\end{center}
\end{figure}

The charged Higgs pair production for the lightest charged Higgs boson $h^\pm_1$ is one of the most interesting production processes. However, for triplet like lightest charged Higgs boson and singlet like $a_1$ the cross-section is very low, as $a_1$ does not couple to the fermions, and the diagram with $h_{125}$ in the propagator is subdominant. The reason is that the coupling $g_{h_1^\pm h_1^\mp h_1}$ of a totally doublet scalar Higgs boson with two totally triplet charged Higgs bosons is prevented by gauge invariance. The triplet charged Higgs pair production is more suppressed  than the single triplet-like charged Higgs production via a doublet-like neutral Higgs boson. In that case pair production cross-section via off-shell doublet type neutral Higgs mediation ($h_{125}$) in s-channel via gluon-gluon fusion is below $\mathcal{O}(10^{-6})$ fb. The coupling of a pair of $h_1^\pm$ to the $Z$ and the $\gamma$ bosons is shown in Figure~\ref{ZphoHpmHpm}(a) as a function of the doublet fraction. The coupling $g_{h_1^\pm h_1^\mp \gamma}$ is independent of the structure of $h_1^\pm$ as it should be because of the $U(1)_{\rm{em}}$ symmetry. In fact the value of this coupling is just the value of the electric charge. Conversely, the coupling of the $Z$ boson to a pair of charged Higgs depends on the structure of the charged Higgs. When the charged Higgs is totally doublet, its coupling approaches the MSSM value $\frac{g_L}{2}\frac{\cos\,2\theta_w}{\cos\theta_w}$. If the charged Higgs is totally triplet the value of the coupling is $g_L\cos\theta_w$, the same of the $W^\pm-W^\mp-Z$ interaction.
In Figure~\ref{ZphoHpmHpm}(b) we show the variation of the cross-sections with respect to the 
 lightest charged Higgs boson mass $m_{h^\pm_1}$. The colour code of the points are as the previous ones. We can see that for triplet-like points with mass around $\sim 100$ GeV the cross-section reach around pb. This large cross-section makes this production a viable channel to be probed at the LHC for the light triplet type charged Higgs boson. 

\subsection{Triplet charged Higgs phenomenology}\label{pheno}
As already pointed out before, the TNMSSM with a $Z_3$ symmetry allows a very light singlet-like pseudoscalar in its spectrum, which turns into a pseudo-NG mode in the limit of small soft parameters $A_i$ \cite{TNMSSM1}. The existence of such a light and still hidden scalar prompts the decay of a light charged Higgs boson $h^\pm_1 \to a_1 W^\pm$. Such decay is only allowed by the mass mixing of the singlet with the doublets or the triplet.  In the extended supersymmetric scenarios with only triplet, one cannot naturally obtain such light triplet-like pseudoscalar and mostly light pseudoscalar mass stays above the lightest charged Higgs mass \cite{pbas3}.  Imposing $Z_3$ symmetry would be impossible due to existence of $\mu$ term, which is necessary for TESSM to satisfy the lightest chargino mass bound \cite{pbas1}. If the lightest charged Higgs boson is pair produced, it can have the following decay topologies
\bea\label{fs1}
pp \to h^\pm_1h^\mp_1 \to  a_1 W^\pm Z W^\mp \nn &\to&  2\tau (2b)+ 2j+ 3\ell \nn +\etmiss \\
   & \to  & 2\tau (2b)+ 4\ell +\etmiss .
\eea
Eq.~\ref{fs1} shows that when one of the charged Higgs bosons decays to $a_1 W^\pm$, which is a signature of the existence of singlet-type pseudoscalar, and the other one decays to $ZW^\pm$, which is the triplet signature.  The tri-lepton and four-lepton backgrounds are generally rather low in SM. In this case they are further tagged with $b$ or $\tau$-jet pair, which make these channels further clean \cite{TNMSSM3, pbac}.

\section{NMSSM}\label{chnmssm}
\begin{figure}
\begin{center}
\includegraphics[width=2.2in,height=3.5in,angle=-90]{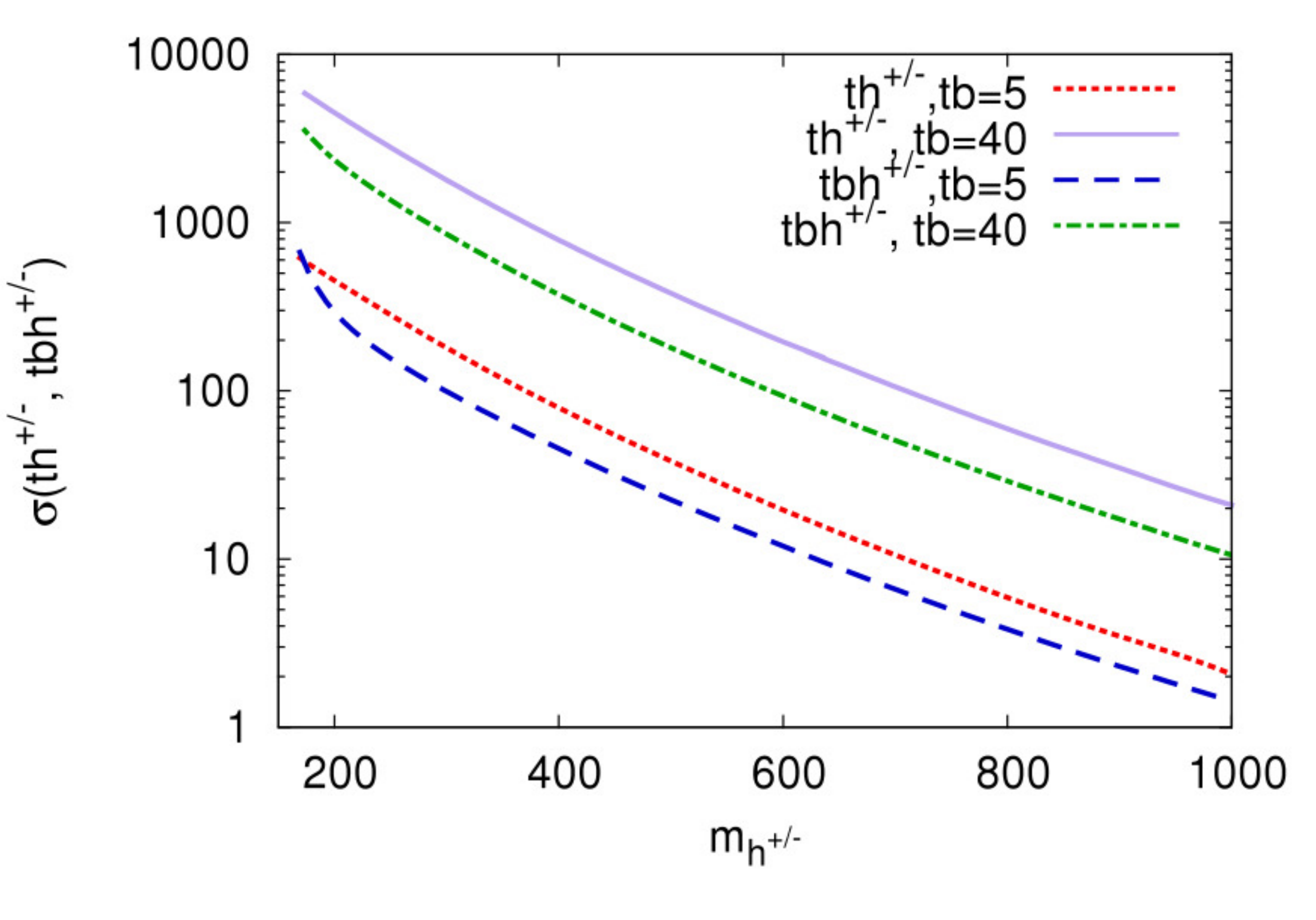}
\caption{Cross-section for $pp \to t h^{\pm}$ and $pp\to tb h^\pm$ vs mass of the charged Higgs boson. The blue, green are
for $tbh^\pm$ and red, violet are for $th^\pm$ production processes at ECM=14 TeV for $\tan{\beta}=5, 40$ respectively (see text). \cite{pbsnkh}.}\label{cross}
\end{center}
\end{figure}

\vspace*{-0.25 cm}
 NMSSM with $Z_3$  symmetry also can have a light pseudoscalar mode
 as pseudo-NG mode. Possibility of such light pseudoscalar gives rise to the decay of the doublet like charged Higgs to $a, W^\pm$, where $a$ is the pseudoscalar in NMSSM \cite{chargedhiggs-pheno}. Unlike TNSSM or TESSM, NMSSM has only one charged Higgs boson like in MSSM or 2HDM which is doublet type thus couples to fermions. For the case with $m_{h^{\pm}} > m_t$, the dominant production modes of the charged Higgs boson comes $bg$ fusion \cite{pbsnkh}. In this case 
we produce a single charged Higgs boson in association with a top quark. Its coupling to top and bottom quarks has two parts: one is proportional to $m_t \cot{\beta}$ and the other part is proportional to $m_b \tan{\beta}$. This feature makes the top (or bottom) mediated production modes highly  $\tan \beta$ 
dependent as can be seen from Fig.~\ref{cross}.

The presence of a light pseudoscalar gives $b$- or $\tau$-rich final state which helps to avoid the SM backgrounds. 
We investigated the $1b + 2\tau + 2\ell +\ptmiss$, $1b + 2\tau + 2j + 1\ell + \ptmiss$ and $3b + 2\ell + \ptmiss$ final states resulting from $W^{\pm}$ decay modes. A detailed cut-based analysis shows that such  scenarios can be probed with the data of as little as $\sim 10$ fb$^{-1}$ of integrated luminosity at the LHC with 13 TeV and 14 TeV center-of-mass energy \cite{pbsnkh}.  
\section{Discussion and conclusions}\label{dis}
 We focus our attention on a typical mass spectrum with a doublet-like CP-even Higgs boson around 125 GeV,  a light triplet-like charged Higgs boson and a light singlet-like pseudoscalar. The existence of light singlet-like pseudoscalar and triplet-like charged Higgs  boson enrich the phenomenology at the LHC and at future colliders. In general we expect to have mixing between doublet and triplet type charged Higgs bosons. We find that in the decoupling limit, $\lambda_T \simeq 0$, one should expect two triplet-like and one doublet-like massive charged Higgs bosons. However since the Goldstone boson is a linear combination which includes a triplet contribution $\sim {v_T}/{v}$ \cite{TNMSSM3}, one of the massive eigenstates triplet cannot be $100\%$ triplet-like. 

Recent searches by both CMS \cite{ChCMS} and ATLAS  \cite{ChATLAS} are conducted for a 
charged Higgs mainly of doublet-type and coupled to fermions. For this reason such a state can be produced in association with the top
quark and can decay to $\tau\nu$. Clearly, these searches have to be reinvestigated in order to probe the possibility of triplet representations of $SU(2)$ in the Higgs sector.
The breaking of the custodial symmetry via a non-zero triplet vev generates $h_i^\pm-W^\mp-Z$ vertex 
at the tree-level in TNMSSM. This leads to the vector boson fusion channel for the charged Higgs boson, which is not present in the MSSM or the 2HDM at the tree-level.  On top of that the $Z_3$ symmetric 
superpotential of TNMSSM has a  light pseudoscalar $a_1$ as a pseudo NG mode of a global $U(1)$ symmetry, known as the "$R$-axion" in the literature. However the later can also be  found in the context of the $Z_3$ symmetric NMSSM. In this case the light charged Higgs boson can decay to $a_1 W^\pm$ \cite{chargedhiggs-pheno, pbsnkh} just like in the TNMSSM. In the context of a CP-violating MSSM, such modes can arise due to the possibility of a light Higgs boson $h_1$ and of CP-violating interactions. A charged Higgs boson can decay to $h_1 W^\pm$ \cite{CPVMSSM}, just as in our case. Therefore, one of the challenges at the LHC will be to distinguish among such models, once such a mode is discovered.

Triplet charged Higgs bosons with $Y=0$, however, have some distinctive features because they do not couple to the fermions, while the fusion channel $ZW^\pm$ is allowed.  The phenomenology of such triplet-like charged Higgs boson has already been studied in the context of TESSM \cite{pbas3}. Such charged Higgs bosons also affect the predictions of $B$-observables \cite{pbas1, pbas2} for missing  the coupling to fermions and the neutral part does not couple to $Z$ boson.  However in TESSM, even though the charged Higgs boson decays to $ZW^\pm$ \cite{pbas3}, the possibility of a light pseudoscalar is not so natural \cite{pbas1, pbas2, pbas3}. Indeed, one way to distinguish between the TESSM and the TNMSSM is to exploit the prediction of a light pseudoscalar in the second model, beside the light triplet type charged Higgs boson. 

The triplet type light charged Higgs boson in the TNMSSM is allowed to decay both to $ZW^\pm$ as well as to $a_1 W^\pm$, the former being a feature of the triplet nature of this state, and the latter of the presence of an $R$-axion in the spectrum of the model. Unlike NMSSM, in TNMSSM with a $Z_3$ symmetry the decay $h^\pm_1 \to ZW^\pm$ is possible for a triplet-type light charged Higgs boson. We are currently performing a detailed simulation of both the TESSM and the NMSSM in order to identify specific signatures which can be compared with the TNMSSM \cite{pbac}. In Finding these decay modes can surely be a proof of the existence of both the singlet and the triplet in the mass spectrum which is smoking gun signature for TNMSSM. Distinguishing among the doublet, triplet like charged Higgs boson surely give us a more clues about the EWSB and the role of other Higgs bosons from different $SU(2)$ representations. 
\vspace*{-0.5 cm}
\section*{Acknowledgement}
\vspace*{-0.25 cm}
PB acknowledges INFN travel grant to attend the LHCP 2016 at Lund University, Sweden.

\end{document}